\documentclass[conference]{IEEEtran}
\IEEEoverridecommandlockouts

\usepackage{cite}
\usepackage{amsmath,amssymb,amsfonts}
\usepackage{algorithmic}
\usepackage{graphicx}
\usepackage{textcomp}
\usepackage{xcolor}
\usepackage{array}
\usepackage{booktabs}
\usepackage{multirow}
\usepackage{url}
\usepackage{pgfplots}
\pgfplotsset{compat=1.18}
\usetikzlibrary{patterns,positioning}
\pgfplotsset{
  scaleline/.style={
    width=\columnwidth, height=5.5cm,
    ymajorgrids, grid style={gray!25},
    tick label style={font=\footnotesize},
    label style={font=\small},
    legend style={font=\footnotesize, at={(0.02,0.97)}, anchor=north west, draw=none, fill=white, fill opacity=0.7, text opacity=1},
    mark size=2pt,
    xlabel={Cluster size (nodes)},
  },
}

\begin{document}

\title{Fleet-Scale Pod Deployment with VPC-Native Networking in Managed Kubernetes}

\author{
\IEEEauthorblockN{Sri Saran Balaji Vellore Rajakumar, Jayanth Varavani,
Murat Parlakisik, Pavani Panakanti}
\IEEEauthorblockA{\textit{Amazon Web Services}}
}

\maketitle

\begin{abstract}
Managed Kubernetes pod deployment is often presented as a choice between
VPC-native and overlay networking. Public EKS documentation describes prefix-mode
capacity and configuration, while prior CNI studies primarily measure
steady-state throughput and latency. This paper presents an operational
measurement study of fleet-scale pod deployment across three VPC-native modes on
Amazon EKS: individual secondary-IP allocation, ENI preallocation, and IPv4
prefix delegation, using Cilium and Calico as overlay baselines. We compare
representative mode-specific warm-pool settings, so the results describe
combined operating points. At 400 nodes, the tested three-address secondary-IP
warm pool required 7,403 seconds to place 80,000 pods. With the tested
ENI-preallocation and prefix-delegation settings, the same workload completed
in approximately 495 seconds, moving address provisioning off the pod-creation
critical path. In the reported 400-node runs, prefix delegation achieved
approximately 162 pods/s, within the same observed range as the fastest overlay
baseline and ENI-preallocation runs, while requiring fewer successful EC2
provisioning calls per pod than individual-IP allocation by allocating addresses
in \texttt{/28} blocks. Because each \texttt{/28} contains 16 addresses and
occupies one IPv4 slot, an interface with 30 slots carries 29 prefixes,
representing up to 464 potential pod addresses, compared with approximately
30 IPv4 slots per ENI for individual-IP allocation when kubelet and CNI limits
are configured accordingly. Prefix delegation also has lower
documented network-address usage and retains direct VPC reachability and flow-log
visibility; its trade-off is that \texttt{/28} allocations require contiguous
subnet space. Single-node measurements confirm the density effect directly rather
than from documented limits: under individual secondary-IP allocation the attainable
pod count equalled the node's address supply exactly on four instance types (27, 56,
232 and 232 pods), whereas prefix delegation provided up to sixteen times more
potential pod addresses; on three of the four instance types, workload capacity
stopped pod placement before that address supply was exhausted, leaving 104 to
1,312 addresses unused.
\end{abstract}


\section{Introduction}
\label{sec:introduction}

Linux containers made it practical to package and deploy applications at high
density~\cite{felter2015updated, soltesz2007container}, while Kubernetes
provided the orchestration layer for managing those applications across fleets
~\cite{burns2016borg}. Kubernetes does not implement pod networking itself.
Instead, it delegates connectivity to external plugins through the Container
Network Interface (CNI), so the operator's choice of plugin influences the
cluster's addressing model, data path, and cloud integration
~\cite{minna2021understanding, qi2020assessing}.

As Kubernetes moved onto public clouds through managed offerings such as Amazon
EKS, Google GKE, and Azure AKS~\cite{aws-docs-vpc-cni, gke-networking,
azure-cni}, each provider had to reconcile the orchestrator's flat networking
model with the primitives of the underlying software-defined network. That
reconciliation centers on how pods obtain IP addresses. An \emph{overlay} model
gives each pod an address from a cluster-internal range and tunnels pod-to-pod
traffic over VXLAN or IP-in-IP. This decouples pod addressing from cloud limits
at the cost of encapsulation and additional tunnel processing
\cite{calico2020, kapocius2020performance}. A \emph{VPC-native} model assigns
each pod an address from the VPC address space. With prefix delegation, IPAMD
requests /28 prefixes and EC2 assigns them to ENIs; IPAMD then allocates
individual pod addresses from those prefixes. Pods remain directly reachable through the cloud network and visible
in VPC flow logs. The trade-off is that pod density and deployment behavior are
coupled to cloud interface, address, and control-plane API limits
\cite{aws-docs-vpc-cni, gke-networking}.

That coupling becomes more consequential as clusters grow. Operationally, VPC
CNI address management has evolved from small pools of individual secondary IPs,
to ENI-level preallocation, and then to delegated prefixes. Each step keeps more
capacity ahead of pod demand, but changes the unit in which the cloud control
plane is asked to provide that capacity. On Amazon EKS, the resulting behavior is
shaped by per-instance ENI limits, per-ENI address or prefix limits, account-level
EC2 API limits, and finite IPv4 subnet capacity. Prior CNI studies have focused
primarily on steady-state throughput and latency
\cite{dakic2024cni, budigiri2021network, shah2021quantitative,
qi2020assessing}, while deployment speed has received less attention. In
production, however, autoscaling, rolling upgrades, and recovery from failure
can make the time required to bring thousands of pods to Ready the binding
operational constraint.

This paper presents an operational, fleet-scale measurement on Amazon EKS. We
compare representative mode-specific operating points rather than equal
warm-address budgets, so the results characterize combined trade-offs rather
than isolate allocation granularity. In the tested settings, small
individual-address pools leave allocation on the pod-creation critical path;
ENI preallocation and prefix delegation move more work ahead of pod creation and
reach the overlay range. Prefix delegation adds fewer interface attachments,
higher documented address capacity, and retained direct VPC reachability and
flow-log visibility.

The remainder of this paper is organized as follows.
Section~\ref{sec:background} surveys pod networking models and related
measurement work. Section~\ref{sec:architecture} describes the system
architecture and the interface-attachment call flow for each configuration.
Section~\ref{sec:methodology} presents the measurement methodology.
Section~\ref{sec:results} reports the fleet deployment results across all
configurations and scales. Section~\ref{sec:discussion} synthesizes the
findings into practical configuration guidance. Section~\ref{sec:conclusion}
concludes with future work directions.

\section{Background and Related Work}
\label{sec:background}

Kubernetes delegates pod connectivity to CNI plugins that implement the ADD and
DEL operations for attaching and detaching network interfaces to pod
namespaces~\cite{minna2021understanding}. The two addressing models introduced in
Section~\ref{sec:introduction} differ in where that address comes from: an ENI in
the cloud virtual network~\cite{aws-docs-vpc-cni}, or a cluster-internal pool
whose traffic is encapsulated between hosts~\cite{calico2020, cilium2020}.

Either choice is independent of the forwarding engine, and both models can adopt
an accelerated one. Modern CNIs increasingly
move packet processing from the conventional \texttt{iptables} path to
eBPF programs attached at the traffic-control or eXpress Data Path (XDP)
hooks, which run in the kernel and process packets before, or in place of, the
netfilter path~\cite{hoiland2018express, miano2019creating}. The advantage is
most pronounced for policy and service lookups: an \texttt{iptables} chain is
traversed linearly, so enforcement cost grows with rule count, whereas an eBPF
map supports near-constant-time lookup that scales far better as policy and
service counts grow. The engine is orthogonal to the addressing model, so the same
data path can be paired with either an overlay or a VPC-native address source.

Recent measurement studies have benchmarked CNI plugins on data-path metrics
such as TCP throughput, UDP latency, and HTTP request
rate~\cite{dakic2024cni, budigiri2021network, shah2021quantitative,
qi2020assessing, kapocius2020performance}, and system-tuning studies have
examined how kernel and CNI configuration profiles affect these
metrics~\cite{optimizing2024cni}. These studies primarily evaluate steady-state
performance, whereas this paper measures deployment-time scalability: how fast
a cluster can provision addresses and bring pods to a Ready state under a
fleet-wide burst. Our operational measurements complement this literature by
focusing on the address-provisioning path across the configurations studied
here, rather than introducing a new networking mechanism.

\section{System Architecture}
\label{sec:architecture}

EKS pod networking divides responsibility along a clean boundary: a node-local
\emph{data plane} performs fast IP wiring, while an EKS-managed \emph{control
plane} orchestrates the privileged cloud resources.

\subsection{CNI Components}

In the VPC CNI, the node-local IP Address Management Daemon (IPAMD) maintains a
\emph{warm pool} of pre-allocated IP addresses so that pod scheduling need not
block on synchronous EC2 API calls. On node join, IPAMD attaches ENIs and
pre-allocates secondary IPs or /28 prefixes according to the
\texttt{WARM\_ENI\_TARGET}, \texttt{WARM\_IP\_TARGET},
\texttt{MINIMUM\_IP\_TARGET}, and \texttt{WARM\_PREFIX\_TARGET} knobs. A
background loop, with a 5-second monitor interval, replenishes the pool by
calling \texttt{AssignPrivateIpAddresses} for secondary IPs or delegated
prefixes and creating and attaching an ENI when another interface is needed.
Because the pool is the buffer between pod demand and the cloud API, its size
and refill rate determine whether allocation stays off the pod-creation path.

The overlay CNIs replace that cloud-coupled component with a cluster-internal
one. Both Cilium and Calico draw pod addresses from a CIDR outside the VPC
range, and in both a cluster-level controller delegates a per-node block from
that range: Cilium's operator carves a /24 per node from its cluster pool, while
Calico's IPAM assigns /26 blocks. Because the block arrives when the node joins,
the node-local agent satisfies every subsequent pod request from it, so there is
no per-pod cloud API call to rate-limit and no warm-pool target to size. The
cost moves to the data path: pod addresses are not routable in the VPC, so
cross-node traffic is VXLAN encapsulated and the host performs encapsulation and
decapsulation on the pod's behalf. The two overlays differ in how forwarding is
implemented, with Cilium using eBPF programs and Calico the iptables-mode Felix
dataplane; both retain kube-proxy in our configuration, which equalizes the
service-proxy layer across all five arms.

\subsection{CNI Interface Attachment Call Flow}

\begin{figure}[htb]
\centering
\begin{tikzpicture}[
  lifeline/.style={thick, gray!60},
  actor/.style={draw, fill=gray!12, minimum height=0.5cm, minimum width=1.35cm, font=\scriptsize\bfseries, align=center},
  msg/.style={->, >=stealth, thick},
  rmsg/.style={->, >=stealth, thick, densely dashed},
  lbl/.style={font=\tiny, midway, above},
]
\def\xK{0} \def\xC{2.1} \def\xI{4.2} \def\xE{6.3}
\node[actor] (K) at (\xK,0) {Kubelet};
\node[actor] (C) at (\xC,0) {CNI\\binary};
\node[actor] (I) at (\xI,0) {IPAMD +\\warm pool};
\node[actor] (E) at (\xE,0) {EC2\\API};
\foreach \x in {\xK,\xC,\xI,\xE} \draw[lifeline] (\x,-0.3) -- (\x,-3.6);
\draw[msg] (\xK,-0.7) -- node[lbl]{\textbf{1.} ADD} (\xC,-0.7);
\draw[msg] (\xC,-1.2) -- node[lbl]{\textbf{2.} request IP} (\xI,-1.2);
\draw[rmsg] (\xI,-1.7) -- node[lbl,above]{\textbf{3.} pool hit: assign} (\xC,-1.7);
\draw[msg] (\xI,-2.3) -- node[lbl]{\textbf{4.} miss: assign or create} (\xE,-2.3);
\node[font=\tiny] at ({(\xI+\xE)/2},-2.55) {Create+Attach};
\draw[rmsg] (\xE,-2.85) -- node[lbl,above]{IP(s)} (\xI,-2.85);
\draw[rmsg] (\xI,-3.35) -- node[lbl,above]{\textbf{5.} assign to veth} (\xC,-3.35);
\end{tikzpicture}
\caption{VPC-native CNI ADD call flow; only step~4 calls the EC2 API.}
\label{fig:callflow}
\end{figure}
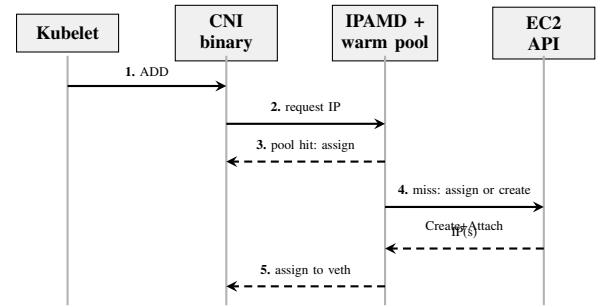

When the Kubernetes scheduler assigns a pod to a node, the kubelet invokes the
CNI plugin's ADD operation. Figure~\ref{fig:callflow} traces this call flow as
a message-sequence ladder; the subsequent path differs fundamentally between
VPC-native and overlay configurations.

The numbered sequence is: (1)~kubelet calls the CNI binary with ADD;
(2)~the CNI binary requests an IP from the local IPAMD; (3)~on a warm-pool hit
the IP is returned immediately with no cloud interaction; and (4)~on a miss,
IPAMD uses the EC2 API. In secondary-IP mode, it assigns individual addresses
with \texttt{AssignPrivateIpAddresses}. In prefix-delegation mode, the same API
can assign a /28 prefix to an existing ENI; when another ENI is needed, IPAMD
creates and attaches it before assigning the prefix. Finally, (5)~the returned
IP is wired to the pod's veth interface. Step~4 is the scalability-critical
path: when many nodes exhaust their warm pools simultaneously, they contend
for shared EC2 API capacity. Prefix delegation softens this contention by
allocating 16 addresses at a time.

The overlay configurations omit step~4 entirely. Their node-local agent draws
the address from the block already delegated to the node and performs only the
local veth wiring of step~5, so no cloud API call appears on the pod-scheduling
path.

\subsection{Configurations Under Test}

\begin{table}[htbp]
\centering
\caption{Networking configurations and warm-pool operating points.}
\label{tab:configs}
\scriptsize
\setlength{\tabcolsep}{2pt}
\begin{tabular}{@{}llcc@{}}
\toprule
\textbf{Config} & \textbf{Model} & \textbf{IPAM setting} & \textbf{Warm IPs} \\
\midrule
VPC-IP        & native  & \texttt{WARM\_IP\_TARGET=3} & $\sim$3   \\
VPC-ENI       & native  & \texttt{WARM\_ENI\_TARGET=3} & $\sim$87  \\
VPC-Prefix    & native  & \texttt{WARM\_PREFIX\_TARGET=10} & $\sim$160 \\
Cilium        & overlay & cluster-pool /24         & local     \\
Calico        & overlay & IPPool /26 blocks        & local     \\
\bottomrule
\end{tabular}
\end{table}

Five configurations are evaluated, spanning both networking models and three
VPC-native address-allocation strategies. Table~\ref{tab:configs} summarizes
their addressing model, IPAM setting, and approximate capacity held ahead of
pod demand.

The three VPC-CNI configurations use representative mode-specific operating
points rather than equal warm-address budgets. We selected three spare
individual addresses, three spare ENIs, and ten spare /28 prefixes to reflect
the allocation units and warm-pool settings operators configure in these modes.
The resulting warm capacities are approximately 3, 87, and 160 pod addresses
per node, respectively, against a workload target of 200 pods per node. Thus,
the experiment evaluates whether each configured allocation unit keeps
replenishment off the pod-creation critical path, rather than comparing equally
sized warm pools or the best possible tuning of each mode.

\textbf{VPC-IP} holds three spare secondary IPs per node. A burst consumes the
pool after three pods, and subsequent requests can wait for replenishment. The
replenishing call is batched rather than per-address: across the 400-node run
0.516 successful calls per pod corresponds to about 1.94 addresses per call, so
pods wait on a shared call rather than each on its own. \textbf{VPC-ENI} maintains
three spare ENIs, so replenishment occurs at interface granularity and most pods
find an address already attached. \textbf{VPC-Prefix} maintains ten /28
prefixes. Each prefix supplies 16 pod addresses, so one allocation serves 16
pods.

The two overlay configurations (\textbf{Cilium}, \textbf{Calico}) have no
equivalent target, since the per-node block described in
Section~\ref{sec:architecture} is delegated at node join and never refilled from
the cloud API. All arms run with kube-proxy enabled (Cilium's
kube-proxy-replacement disabled), equalizing the service-proxy layer.

\section{Methodology}
\label{sec:methodology}

\subsection{Testbed}

The deployment testbed was Amazon EKS version 1.35 in us-west-2 using c6i.4xlarge
instances (16 vCPU, 32 GiB RAM, 8 ENIs with 30 IPv4 addresses each). The pinned
plugin versions are Amazon VPC CNI (\texttt{aws-node}) v1.22.4.eksbuild.3,
Cilium v1.19.5, and Calico v3.29.1. Kube-proxy uses the EKS 1.35 default.
IPAMD warm-pool behavior is controlled only by the environment variables stated
per configuration; all other settings use their defaults. Every arm raises
kubelet's pod cap above the 200 pods per node the workload targets, using
\texttt{--use-max-pods=false} with \texttt{--max-pods} set above 200; the
documented EKS default is 110 and upstream Kubernetes documents 110 pods per node
as a design criterion, so this experiment is deliberately outside that guidance.
All 400-node runs use c6i.4xlarge nodes, which also provide the basis for the
capacity and data-path analysis.
The VPC uses three /17 subnets, one per availability zone, with 32,763
usable IPv4 addresses each after AWS reservations. Overlay configurations use
172.16.0.0/12 as the pod CIDR (1,048,576 addresses).

\subsection{Measurement Protocol}

At each cluster size (10, 50, 100, 200, and 400 nodes), we install the CNI
configuration under test and verify that the cluster is healthy by checking
cross-node pod-to-pod connectivity and ClusterIP service resolution. On the
VPC-native arms we then wait for the IPAMD warm pool to stabilize, so that each
node has finished pre-allocating its configured target and the deployment starts
from a settled state rather than mid-fill. The node counts form a sweep of
distinct fleet operating points.

We then create a single Kubernetes Deployment, in one API call, targeting 200
replicas per node, or 80,000 pods at 400 nodes. Every replica is the same pause
container, requesting 5m CPU and 8\,MiB with no probes. The image and resource
requests are held constant across configurations, making this a controlled
pod-creation workload rather than an application startup benchmark. We use a
watcher on the Deployment to record changes in the ready, creating, pending,
and failed replica counts until the intended target is reached or the ready
count stops advancing for two minutes.

Finally we count the EC2 calls the deployment caused. AWS CloudTrail records the
control-plane API calls made in an account, which lets us attribute address
provisioning to the run rather than infer it. We read the trail's delivered log
objects directly rather than using the \texttt{LookupEvents} API, which is rate
limited and pages fifty events per request: the 400-node VPC-IP window contains
over 800{,}000 EC2 events, which that path cannot retrieve in reasonable time.
From each event we take the action name, the \texttt{errorCode}, and the calling
principal, over the deployment window plus a 180-second buffer to capture
throttled-retry tails that outlast the run. Delivery lag is separate: trail
objects arrive several minutes after the events they contain, so we waited for
delivery to quiesce before reading rather than relying on that buffer.
The error code separates successful allocations from those rejected as throttled, and
the principal separates IPAMD's node-role calls from EKS control-plane interface
management, which shares the same action names. We monitor the six mutating
network actions
(\texttt{Assign}/\texttt{Unassign\-PrivateIpAddresses} and
\texttt{Create}/\texttt{Delete}/\texttt{Attach}/\texttt{Detach\-NetworkInterface}).
Runs are serialized with a quiet period between them and node-group scaling
completes before the window opens, so the counted calls belong to pod creation
and not to node join.

\section{Results}
\label{sec:results}

\subsection{Deployment Throughput at Scale}

\begin{figure}[htb]
\centering
\begin{tikzpicture}
\begin{axis}[scaleline,
  ylabel={Deployment throughput (pods/s)},
  ymin=0, ymax=180,
  xtick={10,50,100,200,400},
  legend style={font=\scriptsize, at={(0.5,-0.25)}, anchor=north, legend columns=3, column sep=4pt, draw=none},
]
\addplot+[thick,mark=*,red!80!black] coordinates {(10,4.3)(50,13.3)(100,11.4)(200,11.3)(400,10.8)};
\addplot+[thick,mark=square*,blue!60!black] coordinates {(10,27.7)(50,136.9)(100,142.8)(200,148.6)(400,161.5)};
\addplot+[thick,mark=triangle*,blue!30!black] coordinates {(10,27.3)(50,135.1)(100,142.8)(200,148.6)(400,161.6)};
\addplot+[thick,mark=pentagon*,teal] coordinates {(10,21.7)(50,109.8)(100,142.8)(200,151.5)(400,161.5)};
\addplot+[thick,mark=star,purple!80!black] coordinates {(10,40.8)(50,135.1)(100,138.8)(200,146.5)(400,143.5)};
\legend{VPC-IP, VPC-ENI, VPC-Prefix, Cilium, Calico}
\end{axis}
\end{tikzpicture}
\caption{Deployment throughput; the 400-node VPC-ENI point uses two runs and
VPC-IP one, while the other 400-node points are repeated-run summaries. All
smaller-scale points are single runs.}
\label{fig:throughput}
\end{figure}
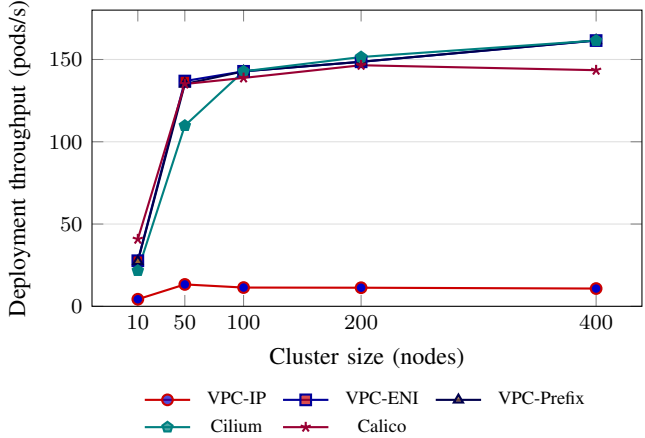

At the tested operating points, deployment throughput separates the
configurations into two regimes that differ by about an order of magnitude
(Fig.~\ref{fig:throughput}). VPC-IP plateaus early and remains near a low rate as
the cluster grows, while the other configurations reach a much higher band at
the larger scale points. Because the separation occurs within the VPC-native
configurations, it is not a simple distinction between VPC-native and overlay
addressing. The rates track whether an address is available locally when a pod
is scheduled under the selected warm-pool settings.

VPC-IP's behavior follows from the \emph{synchronous} EC2 dependency of a
depleted warm pool. Once its $\sim$3 spare IPs per node are consumed, each
further pod waits for an EC2 call before it is wired. The evolved configurations avoid this blocking through different mechanisms.
At the larger scale points, VPC-ENI, VPC-Prefix, and both overlays fall in the
same broad throughput range, although their individual
curves are not identical. VPC-ENI pre-allocates enough IPs per node that
most pods start from the local pool with no EC2 call on the critical path.
Prefix delegation achieves the same effect by amortizing 16 IPs per call.
Overlays bypass the EC2 API entirely by allocating from a local CIDR pool.

\begin{table}[htbp]
\centering
\caption{Repeated runs at 400 nodes. ``$\pm$'' is the sample standard deviation
across runs. Pods/s is the mean of the per-run rates, not 80{,}000 divided by the
mean time; the two differ where variance is high. A coefficient of variation is
reported only where three or more runs are available.}
\label{tab:variance}
\scriptsize
\setlength{\tabcolsep}{3pt}
\begin{tabular}{@{}lcccc@{}}
\toprule
\textbf{Config} & \textbf{Runs} & \textbf{Mean Time (s)} & \textbf{CV} & \textbf{Pods/s} \\
\midrule
VPC-ENI       & 2 & 495.5 $\pm$ 0.7 & ---    & 161.5 \\
VPC-Prefix    & 5 & 495 $\pm$ 1   & 0.2\%  & 161.6 \\
Cilium        & 5 & 495 $\pm$ 1   & 0.2\%  & 161.5 \\
Calico        & 5 & 562 $\pm$ 62  & 10.9\% & 143.5 \\
\bottomrule
\end{tabular}
\end{table}

Repeating the 400-node point separates the configurations by reproducibility as
well as by rate (Table~\ref{tab:variance}). VPC-Prefix and Cilium are the most
predictable, each with a 0.2\% coefficient of variation across five runs, and
their means differ by less than a second. On the retained testbed, the two
deployment rates fall in the same observed range; this comparison is
descriptive rather than a statistical equivalence test. Calico is slower and
more variable at 10.9\%. Every run in the table placed all 80{,}000 pods.

The per-second progress traces provide additional context for these aggregates.
VPC-IP admits pods at a constant low rate for the entire run rather than
slowing as the fleet grows. The other configurations instead admit pods at the
rate the scheduler and kubelet can retire them.

\subsection{Deployment Time at Maximum Scale}

\begin{figure}[htb]
\centering
\begin{tikzpicture}
\begin{axis}[scaleline,
  ylabel={Time to deploy all pods (seconds)},
  ymin=30, ymax=10000, ymode=log, log basis y=10,
  xtick={10,50,100,200,400},
  legend style={font=\scriptsize, at={(0.5,-0.25)}, anchor=north, legend columns=3, column sep=4pt, draw=none},
]
\addplot+[thick,mark=*,red!80!black] coordinates {(10,460)(50,750)(100,1750)(200,3511)(400,7403)};
\addplot+[thick,mark=square*,blue!60!black] coordinates {(10,72)(50,73)(100,140)(200,269)(400,496)};
\addplot+[thick,mark=triangle*,blue!30!black] coordinates {(10,73)(50,74)(100,140)(200,269)(400,495)};
\addplot+[thick,mark=pentagon*,teal] coordinates {(10,92)(50,91)(100,140)(200,264)(400,495)};
\addplot+[thick,mark=star,purple!80!black] coordinates {(10,49)(50,74)(100,144)(200,273)(400,562)};
\legend{VPC-IP, VPC-ENI, VPC-Prefix, Cilium, Calico}
\end{axis}
\end{tikzpicture}
\caption{Deployment time (log scale); the 400-node VPC-ENI point uses two runs and
VPC-IP one, while the other 400-node points are repeated-run summaries. All
smaller-scale points are single runs.}
\label{fig:time}
\end{figure}
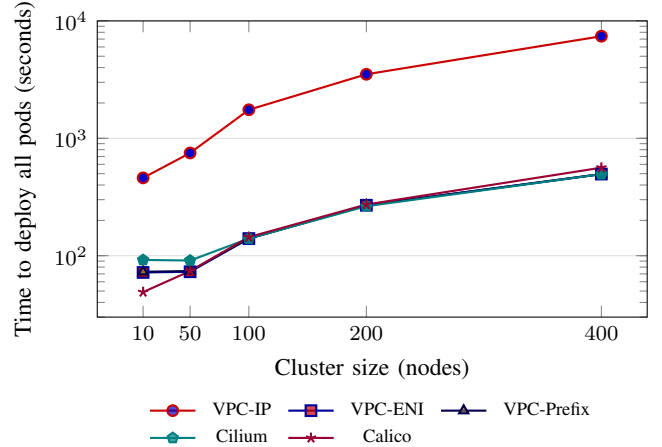

At the largest cluster size, VPC-IP takes just over two hours to place 80,000
pods, while VPC-ENI, VPC-Prefix, and Cilium finish between 493 and 496 seconds
(Fig.~\ref{fig:time}). The pre-allocated arms remain close across the scale
points, so the observed gap is associated with the per-IP allocation strategy
under the tested warm-pool settings rather than with VPC-native addressing in
general.

\subsection{EC2 API Provisioning Work}

\begin{figure}[htb]
\centering
\begin{tikzpicture}
\begin{axis}[
  ybar, bar width=9pt, width=0.95\columnwidth, height=5.0cm,
  ymajorgrids, grid style={gray!25},
  tick label style={font=\footnotesize},
  label style={font=\small},
  symbolic x coords={VPC-IP,VPC-ENI,VPC-Prefix,Cilium,Calico},
  xtick=data, x tick label style={rotate=30,anchor=east,font=\scriptsize},
  ylabel={Successful EC2 provisioning calls per pod},
  ymin=0, ymax=0.60,
  nodes near coords, nodes near coords style={font=\tiny,/pgf/number format/fixed,/pgf/number format/precision=3},
]
\addplot+[fill=red!55,draw=red!70!black] coordinates
  {(VPC-IP,0.516) (VPC-ENI,0.044) (VPC-Prefix,0.014) (Cilium,0) (Calico,0)};
\end{axis}
\end{tikzpicture}
\caption{Successful EC2 provisioning calls per pod fall as the allocation unit grows.}
\label{fig:api}
\end{figure}
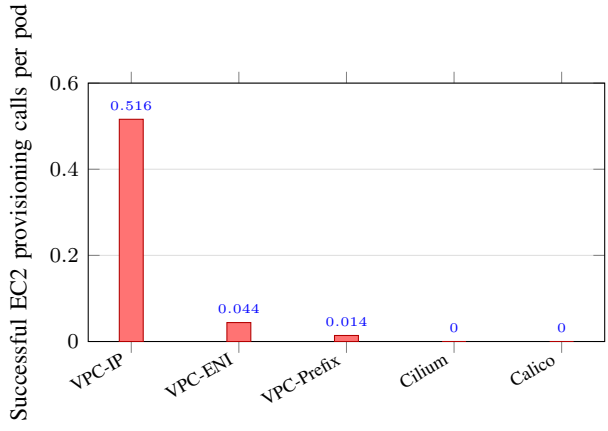

Figure~\ref{fig:api} reports successful EC2 provisioning calls, excluding calls
rejected as throttled. We count calls within the deployment window, with
node-group scaling completed beforehand so that node-join calls fall outside it,
and we classify each event by its CloudTrail \texttt{errorCode}. Overlay
configurations generate no calls for pod IP assignment, as expected.

In these runs, the provisioning-work count declines as the allocation unit
grows. VPC-IP needs 0.516 successful calls per pod, because a depleted
three-address pool leaves nearly every pod waiting on its own allocation.
VPC-Prefix needs 0.014, a factor of 37 fewer, because a single \texttt{/28}
covers sixteen pods. VPC-ENI sits between them at 0.044.
This ordering matches the deployment-time ordering in
Fig.~\ref{fig:time}. Raw CloudTrail counts include throttled retries,
which can dominate under contention. We therefore report successful EC2
provisioning calls separately to isolate provisioning work from retry volume.

\begin{table}[htbp]
\centering
\caption{Documented capacity and network address usage (NAU) accounting for the native modes. These are capacity calculations, not additional density measurements.}
\label{tab:density}
\scriptsize
\setlength{\tabcolsep}{2pt}
\begin{tabular}{@{}lcccc@{}}
\toprule
\textbf{Mode} & \textbf{Unit} & \textbf{Per-ENI limit} &
\textbf{Capacity implication} & \textbf{NAU} \\
\midrule
Secondary-IP & 1 IPv4 address & 30 slots & 232/node & 240 \\
Prefix delegation & /28 (16 addresses) & 29 prefixes &
464/ENI & 31 \\
\bottomrule
\end{tabular}
\end{table}

Table~\ref{tab:density} translates the documented interface and address limits
into the capacity comparison used by this paper
\cite{aws-docs-vpc-cni,aws-prefix-best-practices,aws-vpc-quotas}. A delegated
\texttt{/28} contains 16 IPv4 addresses, and a prefix occupies one of the
interface's IPv4 slots, so an interface with 30 slots carries 29 prefixes and
provides 464 pod addresses when kubelet and CNI pod limits are configured to use
them. In secondary-IP mode, the c6i.4xlarge
instance limit leaves 232 usable pod addresses per node across
its eight ENIs; the documented max-pods figure of 234 for this instance is
$8\times29+2$, where the two additional pods run on the host network and hold no
node address. The same documented accounting assigns 31 NAU to a
prefix-backed ENI and 240 NAU to the secondary-IP configuration. These figures
provide documented context for the density and address-pressure trade-off, not
a separate density benchmark. The rows use different scopes: secondary-IP is
node-level for \texttt{c6i.4xlarge}, while prefix delegation is per-ENI.

\subsection{Attainable Pod Density per Node}
\label{sec:density}

The deployment-time results above hold pods per node fixed at 200. A separate question is
whether the pod address supply limits density at all, and whether prefix delegation changes
that. We ramped pods onto a single node in batches on four instance types in each addressing
mode, and recorded both the pods placed and the addresses left unused, as shown in
Table~\ref{tab:densitymeas}. Density is counted as pods holding a routable VPC address; pods on the host
network, such as the CNI agent and kube-proxy, hold no node address and are excluded.

The supply itself is arithmetic. Writing $E$ for interfaces per node and $I$ for addresses per
interface, secondary-IP allocation supplies $E(I-1)$ addresses, one per slot after the primary.
Prefix delegation supplies $E(I-1)\times16$, because a \texttt{/28} occupies one slot and
carries sixteen addresses, so the mode multiplies the supply by exactly sixteen on every
instance type. IPAMD reports the per-interface figure directly and it agrees: on
\texttt{c6i.xlarge}, \texttt{maxIPsPerENI = 224}, which is $(15-1)\times16$, and
$4\times224 = 896$.

Both supplies are net of the primary address on each interface, which is also how the
documented max-pods figure of 234 for \texttt{c6i.4xlarge} arises: it is
$8\times29+2$, the two being host-network pods that hold no node address and are
excluded here.

\begin{table}[t]
\caption{Pods placed per node and addresses left unused, by addressing mode. ``Reached''
is the highest count of pods simultaneously holding a node address across an arm's trials,
and ``Unused'' is the supply less that count. $n$ is the number of included trials.}
\label{tab:densitymeas}
\centering
\scriptsize
\setlength{\tabcolsep}{2pt}
\begin{tabular}{@{}lrrcrrrc@{}}
\toprule
& \multicolumn{3}{c}{\textbf{Secondary-IP}} & \multicolumn{4}{c}{\textbf{Prefix delegation}} \\
\cmidrule(lr){2-4}\cmidrule(lr){5-8}
\textbf{Instance} & \textbf{Reached} & \textbf{Supply} & \textbf{$n$} & \textbf{Reached} & \textbf{Supply} & \textbf{Unused} & \textbf{$n$} \\
\midrule
\texttt{c6i.large}   &     27 &   27 & 3 &  328 &  432 & 104 & 3 \\
\texttt{c6i.xlarge}  &     56 &   56 & 2 &  718 &  896 & 178 & 3 \\
\texttt{c6i.4xlarge} &    232 &  232 & 3 & 2400 & 3712 & 1312 & 2 \\
\texttt{c6i.8xlarge} &    232 &  232 & 3 & 3712 & 3712 &   0 & 2 \\
\bottomrule
\end{tabular}
\end{table}

Under secondary-IP allocation the address supply is the binding constraint, and it binds
exactly: every trial placed $27 = 3\times9$, $56 = 4\times14$ or $232 = 8\times29$ pods, and in
each the CNI reached the point of having no address left to give the next pod. One
\texttt{c6i.large} trial addressed 25 workload pods with none left to allocate, so co-resident
system pods, which draw from the same supply, held the other two. Density in this mode is
therefore computable from the interface table before a node is provisioned, and because ENI
preallocation draws from the same per-interface slots the same ceiling applies to it: warm-pool
settings change how quickly addresses become available, not how many exist.

Under prefix delegation the supply stops binding. On three of the four instance types the pod
count stopped short of it, leaving between 104 and 1,312 addresses never assigned.
Across trials the counts fell within 14 pods on \texttt{c6i.large} and six on
\texttt{c6i.xlarge}.

On \texttt{c6i.8xlarge} the supply is measured rather than derived. There 3,711 workload
pods held an address and IPAMD reported all 3,712 assigned with none in cooldown, so a
co-resident pod held the remaining one, and its per-interface figure of 464 is
$(30-1)\times16$, giving $8\times464 = 3712$. That this is the only instance to reach its
supply follows from node capacity and not from the mode: \texttt{c6i.4xlarge} has the same
$8\times30$ interface table and therefore the same 3,712 addresses but half the node, and
the two fall either side of the point at which the raised supply becomes reachable, the
smaller leaving 1,312 unassigned while the larger consumed it entire.

Where the node can take up the additional addresses the density gain equals the supply gain: on
\texttt{c6i.8xlarge} the measured ratio is 16.0, against 10.3 to 12.8 on the three instances
that left addresses unused.

\subsection{Steady Cross-Node Data-Path Throughput}
\label{sec:datapath}

The preceding results measure how fast a fleet can be \emph{deployed}. A separate
question is what bandwidth those pods obtain once running, which matters because
overlay pods do not hold their own routable ENI addresses and so share the host
interface. This data-path comparison covers VPC-Prefix, Cilium, and Calico;
VPC-IP and VPC-ENI are outside its scope. We measured this with a many-to-many
experiment: 200 iperf3 server
pods on one host and 200 client pods on another host in the same availability
zone, each client transferring a fixed byte count to its own paired server. All
transfers were released simultaneously by a shared trigger so that pod-start skew
did not enter the measurement. We used the same pod MTU and concurrency on every
arm, allowing the configurations to be compared under otherwise matched
conditions. We report the mean fleet rate over \emph{saturated} seconds, defined
as those in which at least 90\% of pods were transmitting; this excludes the
ramp-up and drain intervals, where partial per-pod samples inflate the apparent
rate.

\begin{figure}[htb]
\centering
\begin{tikzpicture}
\begin{axis}[scaleline,
  xlabel={Bytes transferred per pod},
  ylabel={Steady fleet throughput (Gbit/s)},
  symbolic x coords={50M,100M,500M,1G,2G,5G},
  xtick=data,
  ymin=12.28, ymax=12.58,
  ytick={12.30,12.35,12.40,12.45,12.50,12.55},
  yticklabel style={/pgf/number format/fixed, /pgf/number format/precision=2},
  legend style={font=\scriptsize, at={(0.5,-0.28)}, anchor=north, legend columns=3, column sep=4pt, draw=none},
]
\addplot+[thick,mark=square*,blue!60!black] coordinates
  {(50M,12.412)(100M,12.419)(500M,12.406)(1G,12.407)(2G,12.408)(5G,12.408)};
\addplot+[thick,mark=pentagon*,teal] coordinates
  {(50M,12.498)(100M,12.428)(500M,12.331)(1G,12.329)(2G,12.335)(5G,12.337)};
\addplot+[thick,mark=star,purple!80!black] coordinates
  {(50M,12.362)(100M,12.337)(500M,12.338)(1G,12.338)(2G,12.340)(5G,12.338)};
\addplot[dashed,gray,thick,forget plot] coordinates {(50M,12.5)(5G,12.5)};
\node[anchor=south east,font=\scriptsize,gray] at (axis cs:5G,12.5) {burst ceiling 12.5};
\legend{VPC-Prefix (native), Cilium, Calico}
\end{axis}
\end{tikzpicture}
\caption{Fleet throughput for the three measured arms; all reach the same
host-interface burst rate under the tested conditions. \texttt{c6i.4xlarge} is
documented at 6.25\,Gbit/s baseline and 12.5\,Gbit/s burst.}
\label{fig:datapath}
\end{figure}
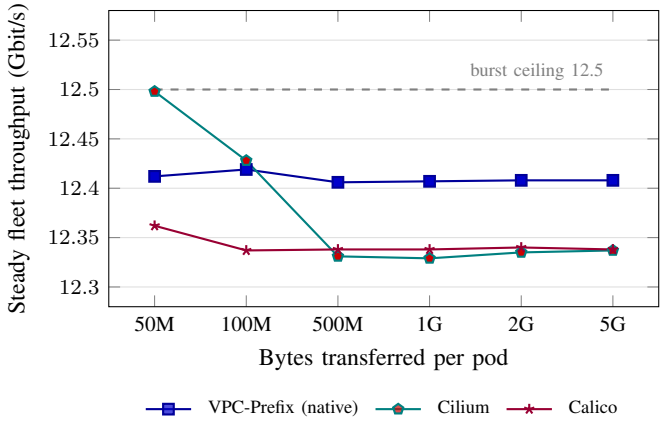

All three configurations converge on the host interface limit rather than on
any CNI-specific bound (Fig.~\ref{fig:datapath}), each reaching more than 98\%
of the instance type's documented 12.5\,Gbit/s burst rate for transfer sizes
above 500\,MB. The spread across the three arms is small enough that the figure's
vertical axis must be truncated to show it at all; on a zero-based scale the
curves would be indistinguishable. The host interface, rather than the CNI,
sets the aggregate bandwidth available to the two-node experiment. Adding pods
subdivides that budget rather than increasing it.

The data-path result therefore does not rank the configurations by throughput.
It shows that prefix delegation retains the same saturated host-interface
behavior as the overlay baselines under the tested conditions, while preserving
VPC-routable pod addresses and flow-log visibility. Any differences at individual
transfer sizes are small relative to the common interface ceiling and are not
the basis for the deployment recommendation.

\section{Discussion}
\label{sec:discussion}

\subsection{Allocation Granularity and the Pod-Creation Path}

The three VPC-native arms combine different allocation units with different
warm-pool operating points. In these measurements, those choices shape both the
deployment and provisioning-work results. One secondary IP serves one pod, one
interface serves dozens, and one delegated prefix serves sixteen. VPC-IP holds
a surplus of three addresses, so its pool is consumed after three pods and
nearly every subsequent pod waits on its own synchronous call; it stays nearly
flat as the fleet grows. VPC-ENI and VPC-Prefix add capacity in units large
enough that most pods find an address already present, and both land in the same
throughput range as the overlay configurations, which draw from a node-local
block and never call the cloud API at all.

The successful-call counts of Fig.~\ref{fig:api} show the same ordering: in these
runs, provisioning work per pod falls as the tested allocation unit becomes
coarser. Together, the results suggest that the practical question is whether
warm-pool replenishment stays ahead of pod demand. A pool need not hold the
entire burst if its refill unit outpaces pod creation for the workload.

\subsection{Overlay Trade-offs at Scale}
\label{sec:overlay}

Overlay configurations achieve consistent throughput without warm-pool tuning,
but they do not remove the host-interface constraint. In the saturated
same-availability-zone experiment, overlay and VPC-native traffic all reached
the same aggregate interface burst rate. As pod density increases, each pod
receives a smaller share of that host budget, so bandwidth-sensitive
workloads still require capacity planning at the instance level.

The important distinction is integration. Overlay pods use cluster-internal
addresses and are not directly visible as pod addresses in VPC flow logs.
Encapsulation and tunnel management are reasonable trade-offs when that
integration is not required. They are less suitable when workloads or operators
depend on direct VPC reachability and VPC-level observability.

\subsection{Why Prefix Delegation Rather Than Interface Preallocation}
\label{sec:whypd}

VPC-ENI and VPC-Prefix both keep allocation off the pod-creation path, so
deployment speed alone does not separate them. In the configurations studied
here, their relevant distinction is the address supply each strategy can reach
and what it spends to get there. The
documented comparison in Table~\ref{tab:density} makes the distinction explicit.
A secondary-IP slot supplies one pod address, whereas a \texttt{/28} supplies
sixteen. On a c6i.4xlarge node, secondary-IP mode therefore provides 232 usable
pod addresses across the eight-ENI limit. Prefix delegation can provide 464
addresses on one ENI, since a prefix occupies one of the interface's 30 IPv4
slots and 29 remain, subject to the kubelet, CNI, and subnet configuration. These
are capacity limits, not an additional density experiment, but they identify
which constraint each mode exposes to an operator.

The same asymmetry appears in the resources consumed to reach a given fleet
size. Because one prefix covers sixteen pods, prefix delegation uses fewer
occupied interface slots and fewer attached interfaces than ENI preallocation.
The deployment runs are consistent with this difference: VPC-ENI attached
roughly 2.7 times as many interfaces as VPC-Prefix to place the same 80{,}000
pods. The documented
NAU accounting in Table~\ref{tab:density} captures the corresponding address
pressure, with 31 NAU for a prefix-backed ENI and 240 NAU for the
secondary-IP configuration.

Prefix delegation's cost is subnet capacity rather than speed. Each
\texttt{/28} must be a contiguous, correctly aligned block drawn from the
node's own subnet, so a subnet can hold ample free addresses in aggregate yet
have no allocatable prefix left, and the assignment then fails for want of a
free CIDR block rather than for want of addresses~\cite{aws-prefix-best-practices}.
Two consequences follow for planning. First, the relevant headroom is
per-subnet, not VPC-wide, because a pod can only draw from the subnet its node
sits in; in our testbed this is why nodes had to be spread one subnet per
availability zone per node group rather than concentrated. Second, the warm
prefix target is the dial that sets how much capacity is committed before any
pod schedules, and setting it too high can commit nearly all of a subnet's
usable addresses to warm pools in advance and strand pods that would otherwise
have fit. Reserving CIDR space for prefixes, or dedicating a subnet to
prefix-mode node groups, removes the fragmentation
risk~\cite{aws-prefix-best-practices}.

\subsection{Putting It Together}

The fleet results make the operational evolution concrete. At 400 nodes, the
three-address secondary-IP warm pool required 7403 seconds to place 80{,}000
pods. Increasing the allocation unit to an ENI or a \texttt{/28} prefix moved
the deployment result to approximately 495 seconds, the same range as the
Cilium baseline. The observed gap is consistent with how VPC-native capacity is
replenished under the selected warm-pool settings, rather than with
VPC-native addressing itself.

For this burst, prefix delegation was the strongest tested VPC-native point. It
supplies sixteen addresses per allocation, uses fewer ENI attachments than
interface preallocation, and has the documented density and NAU advantages in
Table~\ref{tab:density}. It retains VPC reachability and flow-log visibility;
the trade-off is aligned, contiguous subnet planning. ENI preallocation remains
useful when an existing design requires individual secondary IPs, at the cost
of more interface attachments and a lower address ceiling. The small
individual-address pool was unsuitable for the burst tested here.

Overlays remain appropriate when direct VPC reachability and pod-level flow-log
visibility are not requirements. They provide a local address pool without
cloud API provisioning, but retain the integration and encapsulation trade-offs
described in Section~\ref{sec:overlay}. The recommendation is therefore
conditional: use prefix delegation when high-density bursts and VPC-native
integration matter, and choose an overlay when those integration requirements
do not justify subnet and interface planning.

\subsection{Limitations}
\label{sec:limitations}

Several factors define the scope of these results. At 400 nodes, VPC-Prefix,
Cilium, and Calico were repeated five times each for the main large-scale
comparison. VPC-IP has one run, while VPC-ENI has two runs. The smaller scale
points are single runs, so the shape of each curve in Fig.~\ref{fig:throughput}
should be read as one observation per point.

All 400-node deployment arms use c6i.4xlarge nodes. The documented density and
NAU values refer to that configuration. More generally, absolute rates can vary
with instance capacity, software versions, subnet layout, and account-level
capacity.

Because warm-pool settings differ by mode, these measurements do not isolate
allocation granularity from preallocated capacity; they characterize combined
operating points.

The density measurements in Section~\ref{sec:density} are reported per node, on one
node per instance type, so they say what a node accepts rather than how a scheduler
distributes pods across a fleet at that density. They also depend on the workload: the
counts follow from this pod's 5\,m CPU and 8\,Mi memory requests, so a different
request would place a different number of pods, and because instance size and the
request-derived bound move together across the four instance types we cannot separate
them from these arms alone. Each node's \texttt{maxPodsPerNode} was raised above its
address supply, to 582, 1{,}046, 3{,}862 and 3{,}862, so that the kubelet pod cap could
not bind before the addresses did; the documented EKS default is 110. The two addressing
modes ran on separate clusters because \texttt{ENABLE\_PREFIX\_DELEGATION} is a
DaemonSet-level setting and therefore cluster-wide, with both clusters on EKS 1.35 and
VPC CNI \texttt{v1.22.4-eksbuild.3} so that the mode comparison is not confounded by
plugin version.

The deployment experiment uses a single bulk Kubernetes Deployment of identical
pause containers targeting 200 pods per node. It isolates fleet-wide pod creation and
address allocation rather than application startup, rolling updates, or mixed
workload behavior. It does not measure per-pod launch latency or application
readiness. The study evaluates IPv4 pod networking; IPv6 prefix delegation is
left to future work.

The data-path experiment measures aggregate same-availability-zone throughput
between one node pair at MTU 8951 and fixed concurrency. It does not cover
per-pod latency, cross-availability-zone traffic, network-policy processing,
service traffic, or application-level performance. The conclusions are
therefore best read as operational guidance for preserving VPC-native
integration while provisioning address capacity ahead of a high-density
deployment burst, rather than as a universal ranking of all CNI configurations
or workload types. The overlay comparison is limited to Cilium and Calico in
VXLAN mode.

\section{Conclusions and Future Directions}
\label{sec:conclusion}

This study follows an operational evolution within VPC-native Kubernetes
networking. At 400 nodes, a three-address secondary-IP warm pool required 7403
seconds to place 80{,}000 pods. Under the reported 400-node configurations,
ENI-preallocation runs and prefix delegation completed the same workload in
approximately 495 seconds, as did Cilium. The observed difference is consistent
with the allocation granularity and warm-pool settings used in these
configurations. A node that receives one address at a time must repeatedly wait
for the cloud control plane, while a node that receives ENIs or prefixes can
satisfy most pod requests locally.

For this burst, prefix delegation was the strongest tested VPC-native point. A
\texttt{/28} supplies sixteen addresses per allocation, and an interface with 30
IPv4 slots carries 29 prefixes for 464 pod addresses. The
\texttt{c6i.4xlarge} secondary-IP configuration provides 232
usable pod addresses per node, while documented NAU accounting is 31 for the
prefix-backed configuration versus 240 for secondary IPs. These are documented
limits and accounting; the deployment measurements associate the larger
allocation units and selected warm-pool settings with keeping provisioning off
the critical path.

The trade-off is subnet planning. Prefixes require aligned contiguous address
blocks in each node subnet, but the measured data paths still reached the same
host-interface ceiling as the overlay baselines, and the pods retained direct
VPC reachability and flow-log visibility. Within this study's scope, the
practical configuration recommendation is to use prefix delegation when
high-density bursts and VPC-native integration are both important; use ENI
preallocation when the existing design requires secondary-IP addressing; and
use an overlay when direct VPC pod integration is not required.

For future work, we plan to evaluate dynamic route management overhead at
scale (how route table size affects convergence time across CNI configurations)
and Big TCP handling comparisons (how each CNI benefits from kernel support for
larger TSO/GRO segments on high-throughput workloads). We will also extend the
measurement to IPv6 prefix delegation, which assigns /80 prefixes and
fundamentally changes the address-scarcity dimension.

\end{document}